# Content Generation for Workforce Training

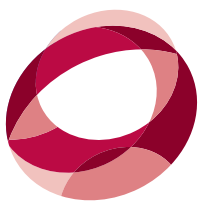

This material is based upon work supported by the National Science Foundation under Grant No. 1734706. Any opinions, findings, and conclusions or recommendations expressed in this material are those of the authors and do not necessarily reflect the views of the National Science Foundation.

# Content Generation for Workforce Training

A Report Based on a CCC Content Generation for Workforce Training Visioning Workshop held March 14-15, 2019, at Georgia Tech, Atlanta, GA.
A Follow-up Workshop by the same name was held July 28, 2019 at the ACM SIGGRAPH Conference in Los Angeles.

**Workshop Chair:**

Holly Rushmeier (Yale University)

**Organizing Committee:**

Kapil Chalil Madathil (Clemson University)

Jessica Hodgins (Carnegie Mellon University)

Beth Mynatt (Georgia Tech)

Tony Derose (Pixar)

Blair Macintyre (Georgia Tech and Mozilla)

Sponsored by

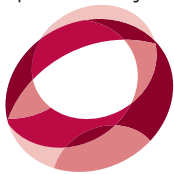

## CCC
Computing Community Consortium
Catalyst

## Summary of Recommendations

Efficient workforce training is needed in today's world in which technology is continually changing the nature of work. Students need to be prepared to enter the workforce. Employees need to become lifelong learners to stay up-to-date in their work and to adapt when job functions are eliminated. The training needs are across all industries – including manufacturing, construction, and healthcare. Computing systems, in particular Virtual/Augmented Reality systems, have been adopted in many training applications and show even more promise in the future. However, there are fundamental limitations in today's systems that limit the domains where computing systems can be applied and the extent to which they can be deployed. These limitations need to be addressed by new computing research. In particular research is needed at multiple levels:

▸ Application Data Collection Level Requiring High Security and Privacy Protections:

- New techniques for capturing job knowledge that integrate sensing systems, natural language processing of existing documents and anthropological methods.

- New techniques for effective evaluation of trainees that integrate sensing systems.

▸ Training Material Authoring Level:

- Understanding level of realism needed for different types of training.

- New representations between 2D images and full 3D scenes.

- New methods to maintain relationship between CAD objects and training systems.

- New methods to organize and share digital assets.

- Machine learning techniques to create training scenarios within specific constraints.

- New systems that allow smooth transitions between fully virtual and fully physical.

- New techniques for authoring systems with a human trainer in the loop.

▸ Software Systems Level:

- Development of new VR/AR authoring systems that focus on training issues such as integrating different types of training materials rather than on computer game design.

- Development of new VR/AR software engines to replace game engines that don't allow for detailed simulations needed by specific training scenarios and interpersonal interaction required in training.

- Development of new VR/AR interaction and visualization techniques directly relevant to training, such as where and how tasks are to be performed.

▸ Hardware Level:

- Efficient mobile computing systems to support AR/VR deployment at low cost.

- Hardware for interaction that is ADA compliant.

- Efficient/secure networking for training of groups.

To accomplish these research goals, a training community needs to be established to do research in end-to-end training systems and to create a community of learning and domain experts available for consulting for in depth computing research on individual system components.

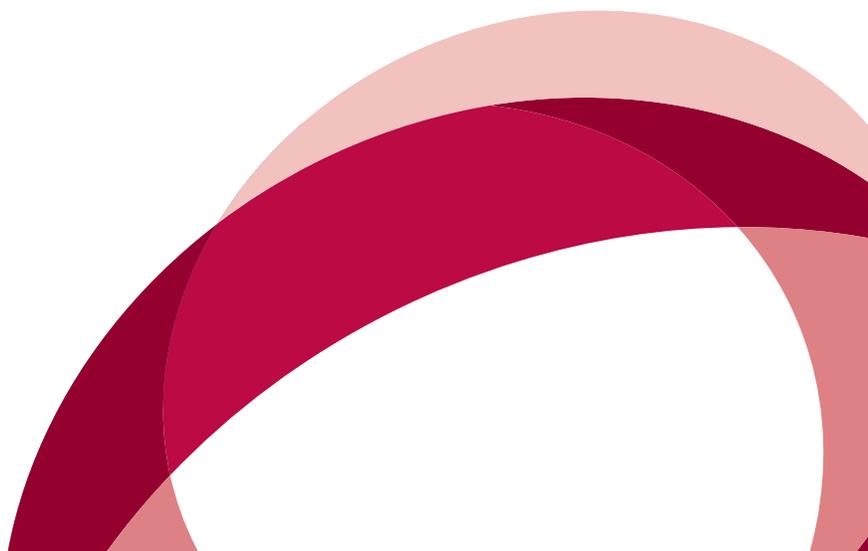

## 1. Introduction

Global competition and the rapid pace of change driven by technological advances are pressuring industries, companies, and individual businesses to pursue solutions for continuous learning for the workforce. Tools and infrastructure are needed to support the continuous development of the US workforce. Individuals entering the workforce require preparation, and they need support to update their knowledge and skills as they advance through their careers. The need for continuous workforce training requires the development of new learning systems. These new learning systems require the invention of new computing technologies to assist in the creation and delivery of content.

This report is based on presentations and discussions at the Computing Community Consortium (CCC) Workshop "Content Generation for Workforce Training" that was held March 14-15, 2019 in Atlanta, Georgia, and a follow-up workshop by the same name held July 28, 2019 at the ACM SIGGRAPH conference in Los Angeles. Here we summarize the context and need for workforce training, challenges in developing new learning systems, and an outline for computing research to meet these challenges.

While growth in the use of information technology has created a demand for education and training in computing and computing related disciplines (including finance), these workshops focused on industrial training needs outside of information technology businesses.

## 2. Context – The Need

Many employers in the field are trying to fill positions that require significant knowledge and training and seeking to update and accelerate skill acquisition of their existing workforce to fill in gaps left by the massive retirement of the Baby Boom generation. To meet these needs, communities are trying to attract new businesses that require a stable pool of qualified labor, leverage education resources to redeploy and retrain underprepared and underutilized individuals, and find fresh ways to recruit and prepare new people into the workforce.

No one solution is universally applicable for creating a more flexible competitive workforce that meets the needs of employers. How do we match the solutions to the various needs? Different industries need unique knowledge and skills. Different sized organizations have varying levels of resources available. Different populations have varying requirements for appropriate instruction.

### 2.1 Varying Knowledge and Skill Requirements

Advanced workforce knowledge and skills are needed across many domains. These domains include goods producing industries such as manufacturing, mining, construction, as well as service industries such as health care and social assistance, leisure and hospitality, and transportation and warehousing. At a high level all of these domains require training on a spectrum including mastering individual manual tasks, trouble shooting and creative problem solving, and effectively working as part of a team. Developing such skills requires experiential training in addition to more traditional readings and lectures. In most cases the experiential learning can be accelerated by controlled exposure to virtual or augmented cases, reducing the learning experience substantially from what would traditionally take months or years to learn "on the job."

However, each workforce domain has specific training requirements that are not readily transferable between applications. Making a measurement on the floor of a manufacturing facility is different from performing a test on a patient in a hospital. Coordinating a team on a construction site is different from organizing a team treating a patient in a hospital emergency room. It is not possible to construct generic sets of assets with simple interfaces for customizing experiences.

## 2.2 Varying Resources

Workforce training is needed for organizations of varying size and resources. Large companies can invest in the development of such computer-enhanced training materials, but they are unlikely to invest in the research in learning and computing necessary to enable the more effective and rapid development of such materials. If given more automated tools that more efficiently leverage the time of in-house training staff, companies will be more likely to train individuals to work in facilities, rather relying on a lower- cost, but better-trained labor force outside of the US.

Many small companies cannot afford to offer training at all, however, communities and larger companies depend on these small businesses. Partnerships have developed between community colleges, federal, state and local governments and businesses to provide training for new and incumbent workers. To facilitate such partnerships, innovative new tools for creating learning systems can allow a relatively small number of training staff to prepare appropriate materials for particular problem domains that can be distributed across the country.

## 2.3 Varying Individual Needs

Individual people do not all start at the same place when joining the workforce. They face different barriers to entry. Special needs and conditions include the transition of veterans back into workforce, people with disabilities who might be underemployed, and aging workers who want to continue to contribute in a meaningful way to the workforce of the future.

Innovative learning systems can be tailored to address these different types of barriers. Veterans that have already acquired experience can be assisted in transferring their knowledge and skills to the civilian workplace. People can have access to training that is adapted to their needs and is not limited by disability. Aging workers can be offered the opportunities to continue to acquire new skills to stay in the workforce and/or make use of tools to transfer their expert knowledge and skills to others when they plan to leave the workforce.

## 3. State of the Art in Research and Practice

Government and industry have long been aware of the need for efficient workforce training and that computing systems can support satisfying this need. In this section we briefly describe long term efforts that have demonstrated the potential of advanced computer supported training and the structures that are already in place to transfer successful training research into practice. The following sections will describe the computing research that is necessary to make order of magnitude advances in using computer systems in training.

## 3.1 Training Research

At the workshops, several research efforts were described illustrating applications of various types of computing systems to a variety of training scenarios. The types of training systems include virtual reality, augmented reality, human-guided systems, and games. The training scenarios include manufacturing, healthcare, education, and construction.

Over the last 25 years virtual reality has been applied to many domains – including applications as diverse as phobia treatment, entertainment, tourism, and architectural design. There has been extensive research in applying virtual reality in training for manufacturing. The immersive nature of VR makes it useful for training tasks such as visual inspection and visual search. Some of these tasks need to be performed in a large scale space (e.g. Fig 1) can't be simply described or done at a workbench, and without VR would require transporting students to

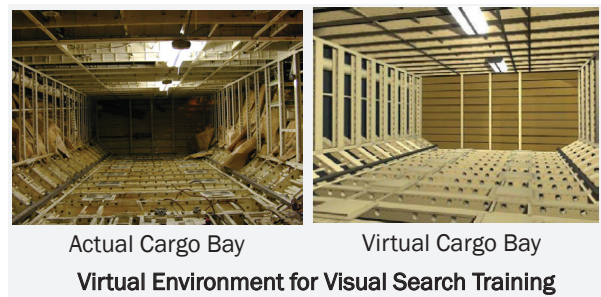

Actual Cargo Bay      Virtual Cargo Bay

**Virtual Environment for Visual Search Training**

*Figure 1: A virtual reality simulation developed in 2004 at Clemson University Center for Workforce Development for training.*

a variety of different physical spaces. Such spaces may not yet exist (they are still be built) or may be expensive to use since normal operations would need to be stopped during training. Some of these tasks require expensive equipment and require introducing trainees to a wide variety of possible problems in a controlled manner – e.g. it could be difficult to maintain a physical library of parts with all of the possible part flaws to be found (e.g. Fig 2).

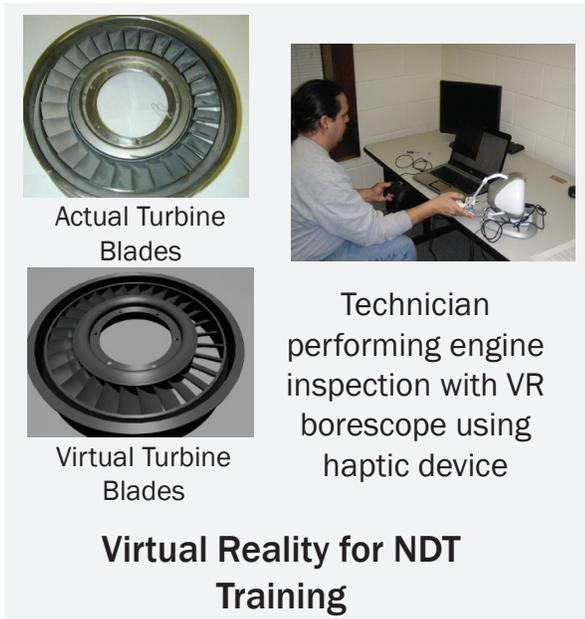

Actual Turbine Blades

Virtual Turbine Blades

Technician performing engine inspection with VR borescope using haptic device

**Virtual Reality for NDT Training**

*Figure 2: A virtual reality simulation developed in 2004 at Clemson University Center for Workforce Development for training.*

The systems are valuable in their ability to immerse trainees in their learning experience. They are also more engaging than passive presentations and videos followed by limited physical interactions.

There are also many training applications of VR (e.g. military missions, civilian police and fire, industries with heavy equipment or that handle hazardous materials) – where the trainee needs to act in dangerous but new, unknown situations (e.g. Fig 3). Creating entirely new environments is costly, but training on the same or similar environments is ineffective because the "new, unknown" factor is lost. An example of VR research is developing efficient systems for rapidly designing new environments with the characteristics needed for a training exercise.

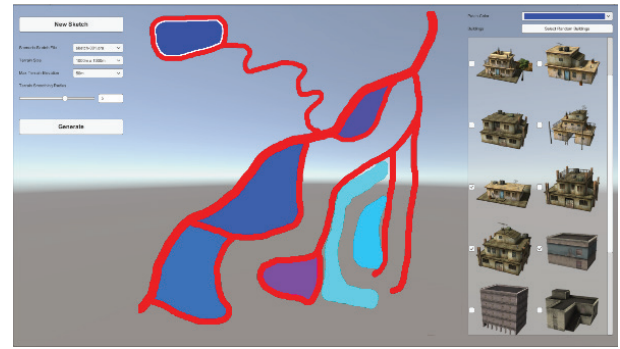

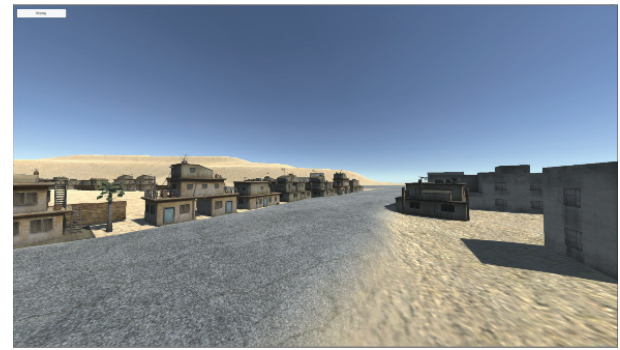

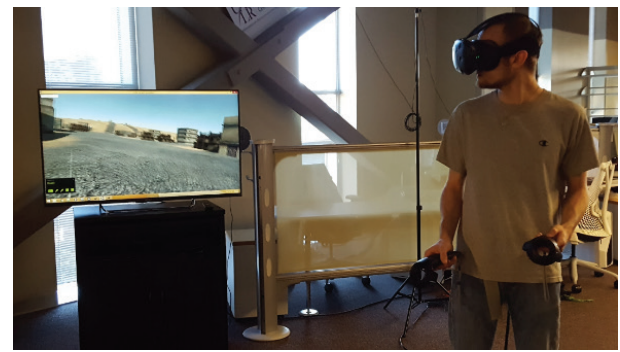

*Figure 3: Scene generator from University of Arkansas Little Rock.*

An interesting development is that virtual reality is not just used to train people for physical interactions. It is also being used for training in education and social skills, such as the TeachLive simulation (marketed commercially by Mursion). Such training providing learners with opportunities to develop focused social management skills by interacting with digital puppets displaying different behaviors controlled by a human in the loop (e.g. Fig 4).

Augmented reality, which combines views of physical and digital objects together, is being used increasingly. This can include manufacturing (e.g. Fig 5), such as adding labels to parts as a mechanic works on them, or imposing computed or measured data such as temperatures and stresses on parts or machines. It also has application in medicine (e.g. Fig 6), to facilitate combining advanced 3D imaging and the physical view of a patient.

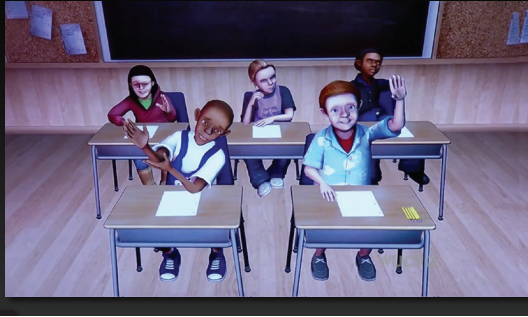

*Figure 4: The TLE TeachLivE™ at the University of Central Florida, is a mixed-reality classroom with simulated students that provides teachers the opportunity to develop their pedagogical practice in a safe environment that doesn't place real students at risk.*

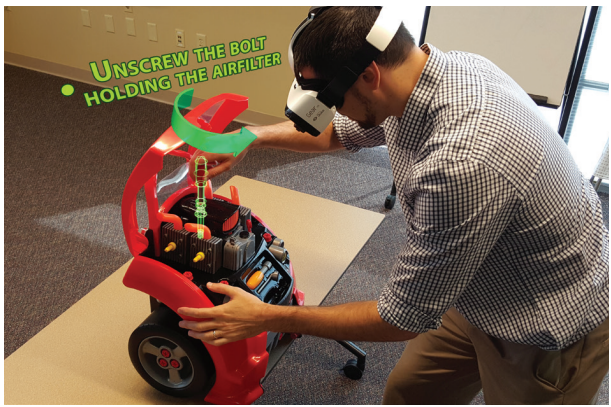

*Figure 5: Labels on a mechanical part, University of Arkansas Little Rock.*

Many systems are being developed that aren't strictly AR or VR, so the term XR (X is something digital) R (reality) is used. An example of a complex visual/physical/interactive system that is being developed is the Physical-Virtual Patient to replace working with live patient "actors" (e.g. Fig 7). A wide range of conditions can be presented. Child patients can be presented, a scenario that can't be simulated with medical actors.

On the other end of the scale of complexity, research on leveraging high quality photography with small changes (such as adding a small animated region for a cinemagraph) to create immersive experiences is being conducted for easing content creation for applications such as construction site inspection (e.g. Fig 8).

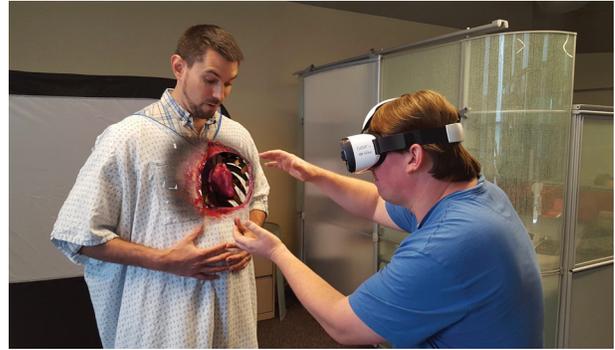

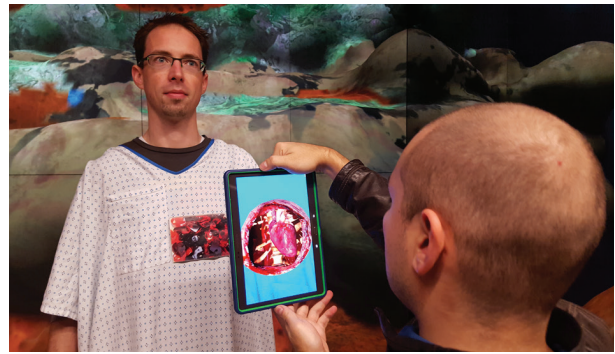

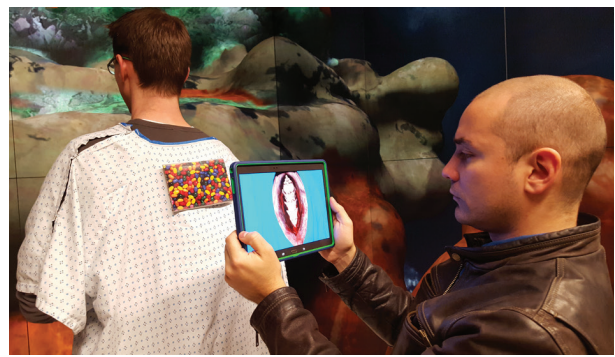

*Figure 6: An augmented reality gown allows a view of data about a patient from various imaging systems, University of Arkansas Little Rock.*

The research has shown that virtual reality training system development is not just presenting visuals or simulations, but must be integrated into frameworks from the educational literature. Virtual reality training needs to be developed in an integrated environment including video and text and trainee assessments.

A substantial amount of proprietary research has been done developing VR systems for training. Large companies contract for these systems from research labs, but then regard them as their competitive advantage. Therefore, many insights are not published, and very little code or data is made open source.

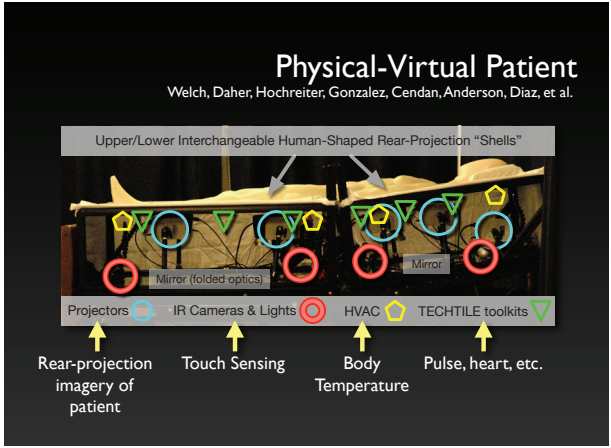

*Figure 7: Virtual Patient design useful for child patients (Photo courtesy Greg Welch).*

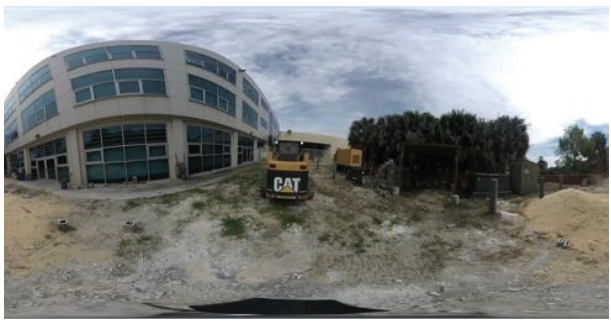

*Figure 8: Cinemagraphs for site inspection (Photo courtesy Eakta Jain).*

## 3.2 Training In Practice

The Federal government has already established Manufacturing Extension Partnerships (MEPs) in every state and Puerto Rico (e.g. Fig 9). In part what these partnerships do is coordinate training both to create jobs (new industries moving in can find trained employees) and retain jobs (retrain people after their previous job function has disappeared for some reason – automation or other shifts in manufacturing). Businesses and industries drive the demand; MEPs work with technical schools to provide the training. Technical schools, rather than the companies themselves, need to provide training since large fractions of the companies involved in manufacturing are very small (i.e. less than 100 employees).

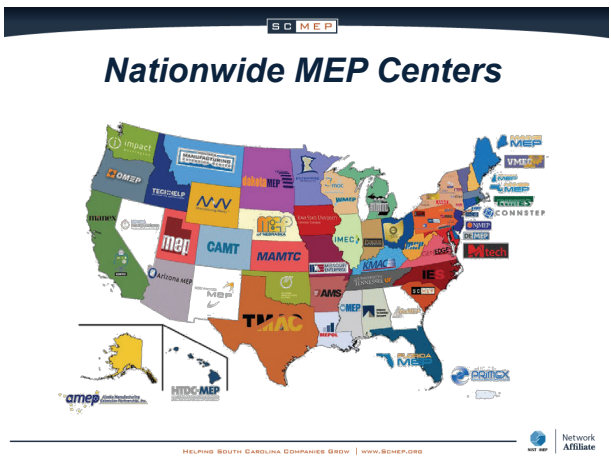

*Figure 9: Centers have been established nationwide for manufacturing training. (Photo courtesy Chuck Spangler)*

Funding is also in place for training for new emerging technologies such as photonic integrated circuits (e.g. Fig 10).

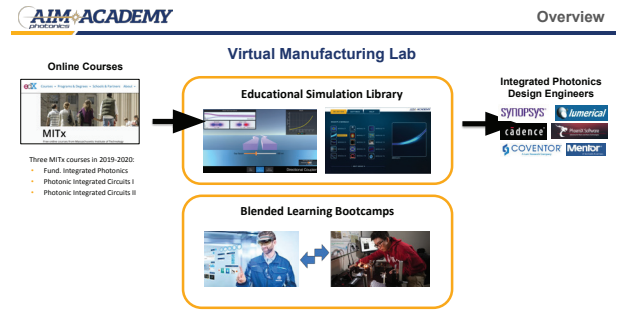

*Figure 10: Virtual Manufacturing Lab (Photo courtesy Erik Verlage).*

Private companies such as SimInsights provide consulting and systems for AR/VR training (e.g. Fig 11). They are also working on products to alleviate the content creation issue such as a system to convert CAD models into interactive objects stored in the cloud for subsequent re-use in AR/VR and simulations.

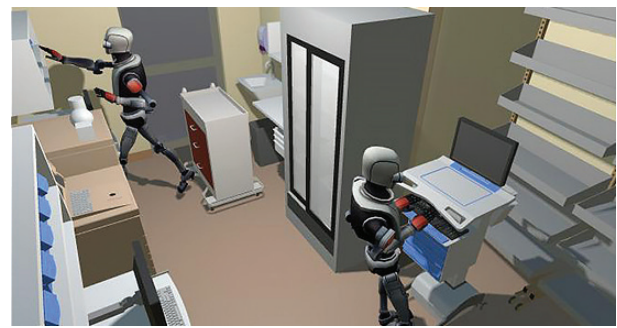

*Figure 11: An environment built with CAD models exported to the cloud (Hypermock, SimInsights).*

Training using the efficiency of computing systems (particularly AR/VR) can address the problem of having an inadequate number of qualified training instructors. This has been recognized by many industries, and training the trainer is a large fraction of VR applications (e.g Fig 12).

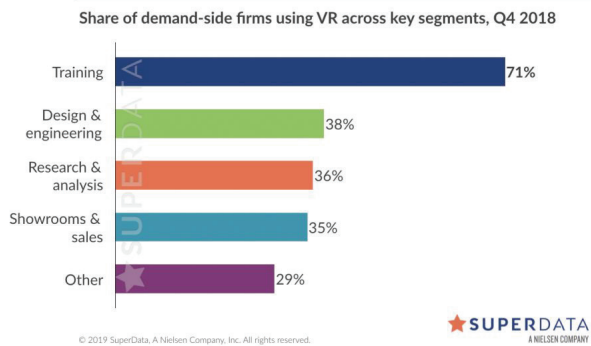

## VR Use By Enterprise

**Share of demand-side firms using VR across key segments, Q4 2018**

| | |
|---|---|
| Training | 71% |
| Design & engineering | 38% |
| Research & analysis | 36% |
| Showrooms & sales | 35% |
| Other | 29% |

*Figure 12: VR applications (Photo courtesy Blair MacIntyre).*

In summary, the promise of computing systems for training has been well established through research and is quickly adopted when available. Structures are in place to transfer technology quickly.

However, there is substantial research documenting the challenges that prevent computing systems, in particular XR systems, from being more broadly used. Some of these are challenges exist in learning systems themselves, which require fresh computing research to advance. Other challenges focus on the need for core training engines that run XR, creating content, and integrating training systems into larger networked environments.

## 4. Challenges in Learning Systems

Learning systems incorporating advanced computing technology have been developed for decades. Flight simulators using analog systems served as the inspiration for experiential learning systems. Today's experiential learning systems involve XR. The systems are not the sole source of training, but are integrated into systems that include manuals, lectures, videos and other traditional teaching techniques. Learning systems using XR have been shown to be successful, but there are still many barriers to widespread use. The challenges here are organized by learning design, implementation and evaluation.

### 4.1 Design of Learning Systems

XR learning designers face several types of challenges. These include knowledge extraction, appropriate application of XR and assessment methods.

**KNOWLEDGE EXTRACTION**

Learning designers need more efficient ways to elicit knowledge from both subject matter experts (SMEs) and existing technical documentation. Innovation is needed for developing best practices and new technologies to support rapid extraction of standards, expert methods, desired learning outcomes, and understanding of local learning needs from a variety of data sources including training manuals and expert interviews. Support is needed to expand the technological resources that assist developers and educators in capturing environmental elements from the workplace that can be annotated to support XR lesson modules. For example, using systematic methods such as evidence-centered design, can help establish re-usable and extensible design resources libraries across multiple workforce fields (for an illustration, see Yarnall & Vahey, 2019).

**APPROPRIATE APPLICATION**

Designers also need guidelines for determining how many XR elements provide "just enough" support to facilitate learning. Designers need for providing guidelines and use cases for defining technical and environmental requirements to develop modular XR units. They also need assistance developing guidelines to define what content to cover with XR and what to assign to other learning modalities. They need help to create blended sequences of XR and hands-on learning experiences. Designers need methods to determine what level of realism is required in XR. They need entry points into existing XR materials so they may rapidly update and locally adapt that content. Guidelines need to be embedded into XR development systems to provide alternate or inclusive immersive experiences for accessibility needs (as required by the Americans with Disabilities Act.)

**ASSESSMENT**

Designers also need tools for embedding assessments and analytic tools into XR learning experiences. Such tools are need support personalization and ongoing evaluation.

Supports are needed to foster early thinking about how to build into XR "safe failure" experiences and collaboration.

In learning design, researchers and developers should prioritize the human element and reach out to workforce communities. Focusing on the human means developing ways to improve design practitioners' understanding of learners' abilities and cultures and the constraints of the different social learning contexts and available technologies. To reduce redundancy, they should consider partnering with workforce content providers, particularly those focused on open educational resources (e.g., Manufacturing, Information Technology, Health Services, etc.).

## 4.2 Implementation of Learning Systems

Trainers and learners need more consistent, reliable channels for accessing XR content. It should be possible to implement training in a modular form, both in the classroom and the workplace. XR should enable virtual experiences for individual and collaborative learning.

Implementation requires giving trainers guidelines on how to integrate XR into their instruction and training environments. They need ways to review and share learner progress through visualizations and dashboards to support instruction using XR tools. Potential solutions might involve developing easily usable guidelines to implement ADA compatible, scalable, device agnostic XR training systems, with the ability to create and refine evidence-based formative and summative assessments in real-time.

Practical guidelines are needed for issues such as how many devices to purchase, how to configure rotations of use, what kinds of shared display modalities can work, etc. These best practices should be published with the provision for community support for iteratively updated refinement of guidelines. Scalability, adaptability, and agnosticism can then evolve with future technologies and requirements.

A primary challenge for implementing XR training involves synthesizing information given by subject matter experts (SMEs) into a tangible training experience. Generally, the first step in creating XR simulations is to meet and brainstorm ideas between the developers, SMEs and instructional designers. The interdisciplinary nature of these meetings provides a challenge in communicating ideas on what the best use of XR is and how much information should be presented to the user. An iterative design process where the SME is continually involved as development continues can help mitigate some issues. It is critical to develop materials in language that is understood by the SME's, trainers and the learners.

## 4.3 Learning Systems Assessment

An essential feature in systems is the incorporation of embedded and extractable assessments in XR. Systems need to incorporate and analyze an adaptive learning model. An adaptive model allows the discovery of trends and gaps and the refinement of the assessment structure.

Further, learning systems must provide methods to evaluate trainers. The evaluation of trainers needs to inform the future training they received to produce more effective delivery in the next iteration.

Essential features of assessment systems include compliance with ADA requirements, and with GDPR (and any future) privacy requirements.

# 5. Computing Research Requirements

Based on the foregoing assessment of the context, state-of-the-art and learning systems challenges, it is clear that many different communities need to work together to address workforce training. Computing alone cannot provide a magic bullet. However, computing research is needed to provide the tools needed by learning systems experts, educators, government and industry to provide training to enable a productive, nimble workforce. We consider these requirements here in sequence from applications-level to systems-level issues.

## 5.1 Application Level – Data Collection

Aspects of systems development that directly involve people are knowledge extraction and evaluation. Research is needed to support and enhance (rather than replace) existing systems for these activities. Because these involve direct interaction and the collection of information about people, these are areas where safety, privacy and consideration of a wide range of abilities are critical.

## KNOWLEDGE EXTRACTION

In knowledge extraction, SMEs are critical. The knowledge SMEs have ranges widely and includes mechanical skills relying on muscle memory, low-level trouble shooting, high-level problem solving and social skills for interacting with coworkers. Methods from anthropology and psychology have been developed to use interview techniques to elicit knowledge from experts. Methods have also been developed to annotate recordings of actions by experts. These methods can be integrated in complementary fashion to knowledge extraction through the use of advanced instrumentation and recording of data. Sensing methods that record as little personal- and individually-identifiable data are needed. Simply using existing system for sensing motions (eye-tracking, motion capture) and audio/visual recording is not enough. Methods are needed to automatically segment this sensed data into micro-tasks. The sensed data then needs to be related to interviews and expert annotations. These different facets can be integrated into machine learning (ML) systems that can assist in rapidly converting interviews and recordings into models of skills that require training.

Expert knowledge is also found in written materials such as manuals and field notes. Methods are needed for extracting training models from these materials, and integrating them with information from interviews and sensing data.

To analyze of the collected data, (ML) techniques should be developed that will efficiently classify the data. Prior user actions/behaviors can be extracted, categorized, and later used for customizing the training experience. In addition to the analysis of user performance, ML and Artificial Intelligence (AI) techniques should be developed to recognize the performance of the users to 1) provide the necessary feedback, and 2) customize the evolution of the training process during the runtime of the application.

## EVALUATION

Evaluation is needed throughout training – both for the trainee and for the trainer. In addition to explicit quizzes or tasks to be passed, sensing methods are needed to assess trainees and adjust the training as it progresses. Trainers also need to be assessed to adjust their approach to future training sessions. Performance could be evaluated from multimodal behavioral data, including gestures, body language, facial expression, eye tracking, as well as physiological data for affect measurement such as galvanic skin response, heart rate, and pupil dilation. While commodity sensors to record such data are available today, as in knowledge extraction open challenges relate to processing this data into user-understandable and system-understandable labels.

As in knowledge extraction, only enough data should be acquired to complete the training. Further, the data that is acquired can be highly personal data that needs to be collected and stored securely.

## 5.2 Material Authoring Level

Creating the digital assets for an XR training system can be enormously time consuming and costly. While the visual, audio and haptic models needed are similar to those used in film and games, training budgets for developing assets in many cases, such as small manufacturers or construction companies, are much smaller. In cases where budgets may be more substantial such as large-scale manufacturing or health care, the assets needed may be qualitatively different, as they require different transitions from virtual to physical or may require a trainer to be actively involved in the training loop. We consider these two cases in turn.

### 5.2.1 Methods for Limited Budgets

In training such as warehouse or construction site inspection, the complexity of the digital assets required to create useful XR training in a straightforward simulation is not cost effective. Several research streams need to be pursued to address this: understanding the level of realism needed for an application, alternative scene representations making use of captured data, importing data from CAD systems in an appropriate form, making use of shared digital assets and employing machine learning to customize and vary environments.

## LEVEL OF REALISM

For different training scenarios – ranging from interacting with digital humans driven by "Wizard of Oz" systems to assembling parts – different levels of realism in presentation are needed. It is a well understood phenomenon that higher levels of accuracy in simulation require more resources, and that many applications can benefit from extremely simple representations. Research is needed in assessing the level of realism that is needed in different types of settings. Furthermore, research is needed to allow the authors of a training system to easily assess the level of realism needed for a particular application.

Level of realism is a critical issue in using digital humans. For 3D virtual avatars, subtle appearance and motion details are perceptually important at close up range for social cues in training scenarios where "reading the other person" is critical. For other training scenarios, it may be equally effective and less cost prohibitive to create lower level of detail or more stylized avatars. We need an understanding of the level of realism that is effective for different training scenarios. The same holds for 3D objects. Background objects may need much lower level of realism than objects that are being manipulated as part of the training. Differing realism requirements may need training situations to be created as videos, user directiable video content (videos++) in addition to computer generated 3D worlds.

Realism is not just an issue of rendering physically accurate form and appearance. Training applications may also require virtual avatars that react to the trainee in ways that vary with the requirements of the scenario. For example, less realistic avatars may react to words spoken by the trainee with dialogue of their own with associated change in facial expression, but more realistic avatars might react to the facial expression of the trainee, or their measured affective state.

### ALTERNATIVE REPRESENTATIONS

Real life settings, ranging from urban settings to unpopulated natural sites, need to be simulated at a high level of visual detail that can be captured in photographs and videos. Training materials can be developed by editing and instrumenting such captured material, as illustrated by the cinemagraph example in Section 3. However editing such materials to appropriately encode training models and instrumenting them is still an open problem.

### IMPORTING CAD SYSTEM DATA

Many training applications involve learning how to use or interact with physical objects that originated in digital form. Methods are needed to smoothly import data from CAD systems, and to even design training in parallel with the initial CAD designs. Furthermore, methods are needed to keep digital objects in synch with their physical counterparts. CAD representations are needed that are instrumented so that a "digital twin" can be maintained for machinery and systems that are used in training. As a physical asset is modified or updated its digital counterpart should be readily updated so that it can continue to be useful in training.

### SHARED DIGITAL ASSETS

Various sites exist online, such as sketchfab.com, for users to exchange 3D data. Guidelines are needed for developing 3D models that are usable in training scenarios with the data that are needed to readily use them. Such models could be used to populate scenes around the machines, building, sites that are critical to the training.

At a high level we need a classification system of the different types of training that need to be performed. Types of systems are not necessarily domain specific, but can vary based on the end result needed for training. For example, different types of training may include learning individual "hands on" tasks, learning tasks that require physical/social interaction with groups, learning inspection tasks, and learning problem solving strategies. A classification of tasks would enable the definition of primitive assets and methods appropriate to the tasks, and definition of best practices. This would allow the development of reconfigurable systems for each class type with only incremental costs, rather than the cost of developed.

Domain-specific templates can be created for common training scenarios (e.g., hospitals, classrooms, factories, labs), which contain the 3D assets of the common objects for a domain (e.g., instruments, test tubes, tables, chairs, sinks for a lab) and their associated interaction modes (e.g., a faucet can be turned on or off). Example layout designs of the training scenario can also be stored to allow quick procedural generation of a layout for training purposes. For example, a virtual lab can be procedurally generated with all the 3D objects put in place. Such a virtual lab can be directly used for training purposes. If desired, the designer can also modify the virtual lab to fit what he or she wants precisely.

#### MACHINE LEARNING

Each 3D digital environment can be time consuming to set up. In general though, many variations of an environment are needed so that a trainee learns the task, rather than the environment. This was illustrated by the terrain variation method illustrated in Figure 3. ML techniques are needed to take a collection of instrumented assets and create training variations of training environments within given constraints. Research is needed both in the techniques and in the specification of appropriate constraints for a particular training scenario.

### 5.2.2 Methods for Virtual/ Physical Transitions and Human-in-the-loop

In many cases training is for learning to act in physical situations where an entire virtual environment cannot simulate the physical situation, but the physical situation needs to be controlled to guide training. An example of this is using a simulator to prepare for child patients, which was discussed earlier in Section 3. In addition to modeling and controlling virtual objects, instrumented physical objects need to be designed. Tools are needed to allow a training system to smoothly guide the trainee from a fully virtual to fully physical experiences.

In addition to creating assets that range from fully virtual to fully tangible, methods are needed to author controls for "human in the loop" avatars. This includes not only puppetry style controls for the avatars, but ways of how they interact with the larger environment, and how they behave when not directly controlled by a trainer.

### 5.3 Software Systems Level

Many current XR systems depend on game engines, such as Unity or Unreal. Game engines have become sophisticated allowing the specification of game logic, providing efficient visual and audio renderings, and game physics engines that simulate physical interactions such as collisions. Game engines provide high performance to users, and have been successfully used in training and serious games.

Because training scenarios are best authored by domain experts, algorithms and techniques need to be developed that support the authors in creating different content types, connecting and switching between different media types (e.g. interactive virtual environment, instructional video, assessment questions), and quickly updating and editing this content. This goal is supported by systems work that considers the authoring tool as a whole, as well as the development of individual components that incorporate user directable "handles." Critically, authors need to be able to direct the trainee's attention during the training process and require a suite of tools to be able to do this. Further development of the assets and specifications of game logic beyond those currently available in game systems is needed to support these goals. An entirely new approach to the structure and presentation of the authoring interface is required to develop training scenarios.

In addition to having content creation interfaces that are not compatible with training development, game engines fundamentally do not allow the incorporation of sufficiently physically accurate simulations that are required in many training situations. As a result, in many applications the software engine to run a training system needs to be written from scratch, rather than leveraging a game engine. Research is needed to define a "training engine" that has the flexibility needed to accommodate physical simulations, and which can be tuned to provide high performance in the scenarios needed.

Interaction techniques for training in AR/VR are different from those that have been devised for game systems. Research is needed on the principled design, development, and evaluation of AR/VR interaction and visualization techniques that specifically address training requirements. These would include showing where tasks are to be performed and showing how tasks are performed. Techniques have been developed and tested for conventional training media, but much less has been done for AR/VR training.

### 5.3 Hardware Systems Level

At the hardware level, there are many requirements that are critical to efficient training, but are not unique to training needs.

As in any application where content creation is time consuming, the authoring of training systems needs to be as independent as possible from specific hardware and operating systems. A specified system should be portable to a wide range of platforms. Systems need to be robust with respect to updated hardware.

Advancements in the hardware that the trainee interacts with directly are needed. Methods are needed to minimize cybersickness. The development of systems that transition between virtual and physical (as discussed in Section 5.2) will also require the development of new interaction hardware. Guidelines are needed to ensure that all interaction hardware is ADA compatible.

In many cases where only a single trainee is involved in training, security and privacy can be maintained by using isolated systems. However, many training scenarios involve groups of people, or a remote coach who needs to communicate with specific trainees. Such networked systems need to be designed for both high performance and security.

In the future training for many applications will be most efficiently and cost effectively distributed through mobile devices. Current devices have independent systems for imaging, vision, graphics, video processing, and acoustics. Along with other XR applications the demands of training applications, will require hardware designs that integrate these systems to achieve the required performance levels.

## 6. The Road Ahead

Support for training has suffered because it isn't a primary consideration in research – the tools and techniques used are inherited from other domains such as games, engineering design, and simulation. To make progress, a research community specially focused on developing computer-based technologies to support the rapid and more efficient development and delivery of content for workforce training is needed. Funding for projects with both breadth and depth are needed.

Interdisciplinary end-to-end applied projects have been successful in demonstrating potential and delivering practical systems. This work needs to be deepened and expanded to find how innovations from across computing (ML, new software systems, new devices) can be applied to training needs. This work also needs to follow an interdisciplinary track so that research projects can address the specific challenges listed in Section 4 and computing requirements addressed in Section 5, such as knowledge capture from experts, code-free authoring of varied environments, new "training" engines to replace game engines are needed. End-to-end domain projects don't provide the resources for the deep dives necessary in these topics to move the area forward to the next level.  Such individual focused projects could benefit from consultation with other training-technology research domain stakeholders, such as learning system researchers, industry training experts, learning scientists, etc. Building a community of registered stake holders who can participate in funded research as consultants would enable individual research projects in HCI, graphics, networking and similar areas to maintain a focus on workforce training.

## 7. Workshop Participants:

Content Generation for Workforce Training Workshop (March 14-15, 2019) in Atlanta, Georgia

| First Name | Last Name | Affiliation |
|---|---|---|
| Paul | Baker | Georgia Institute of Technology / Center for Advanced Communications Policy |
| Jeff | Bertrand | Clemson University |
| Kapil | Chalil Madathil | Clemson University |
| Sandra | Corbett | Computing Research Association |
| Carolina | Cruz-Neira | University of Arkansas Little Rock |
| Khari | Douglas | Computing Community Consortium |
| Austin | Erickson | University of Central Florida |
| Steven | Feiner | Columbia University |
| Maribeth | Gandy | Georgia Tech |
| Patressa | Gardner | Florence-Darlington Technical College |
| Anand | Gramopadhye | Clemson University |
| Rebecca | Hartley | ARM |
| Jessica | Hodgins | Carnegie Mellon University and Facebook |
| Sabrina | Jacob | Computing Research Association (CRA) |
| Alex | Jaeger | University of Arkansas at Little Rock |
| Eakta | Jain | University of Florida |
| Rajesh | Jha | SimInsights |
| Ruth | Kanfer | Georgia Tech |
| Karen | Liu | Georgia Tech |
| Blair | MacIntyre | Mozilla and Georgia Tech |
| Christos | Mousas | Purdue University |
| Elizabeth | Mynatt | Georgia Tech |
| Holly | Rushmeier | Yale University |
| Chuck | Spangler | SCMEP |
| Erik | Verlage | Massachusetts Institute of Technology |
| Tomer | Weiss | University of California, Los Angeles |
| Greg | Welch | University of Central Florida |
| Helen | Wright | Computing Research Association |
| Louise | Yarnall | SRI Education |
| Lap Fai (Craig) | Yu | George Mason University |

Content Generation for Workforce Training (July 28, 2019) at the ACM SIGGRAPH Conference

| First Name | Last Name | Affiliation |
|---|---|---|
| Carolina | Cruz-Neira | University of Arkansas Little Rock |
| Eakta | Jain | University of Florida |
| Rajesh | Jha | SimInsights |
| Holly | Rushmeier | Yale |
| Jerome | Solomon | Cogswell Polytechnical College |
| Greg | Welch | University of Central Florida |
| Louise | Yarnall | SRI Education |

# NOTES

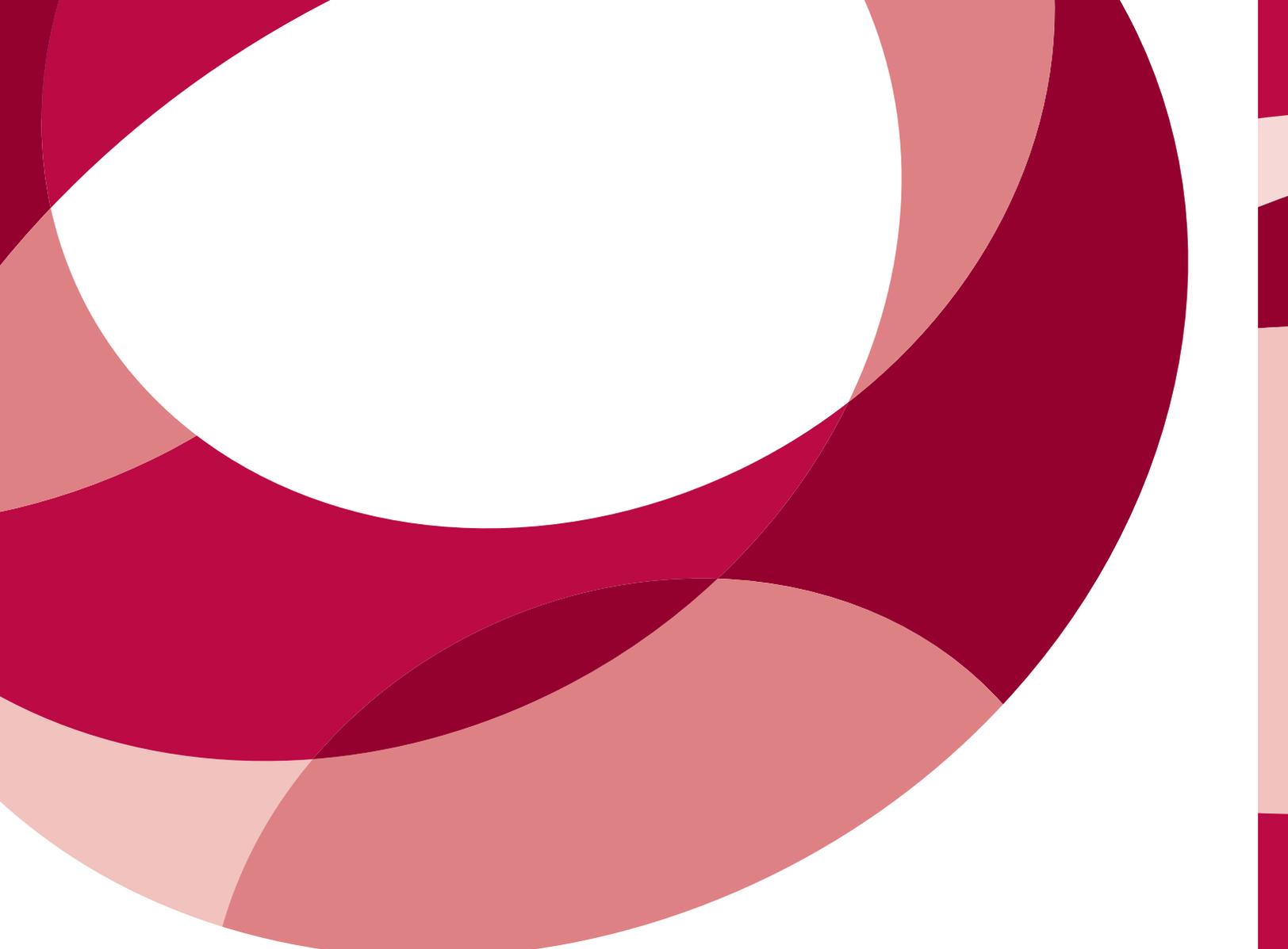

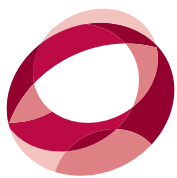

CCC

Computing Community Consortium
Catalyst